\documentclass[%
 aip,
 amsmath,amssymb,
reprint,%
]{revtex4-1}

\usepackage{graphicx}
\usepackage{dcolumn}
\usepackage{bm}

\usepackage[utf8]{inputenc}
\usepackage[T1]{fontenc}
\usepackage{mathptmx}
\usepackage{etoolbox}
\makeatletter\def\@email#1#2{%
 \endgroup
 \patchcmd{\titleblock@produce}
  {\frontmatter@RRAPformat}
  {\frontmatter@RRAPformat{\produce@RRAP{*#1\href{mailto:#2}{#2}}}\frontmatter@RRAPformat}
  {}{}
}%
\makeatother
\begin{document}

\preprint{AIP/123-QED}

\title[Imbibition in conical capillaries]{Thermodynamics and hydrodynamics of spontaneous and forced imbibition in conical capillaries: A theoretical study of conical liquid diode
}
\author{Masao Iwamatsu}
 \email{iwamatm@tcu.ac.jp}
\affiliation{ 
Tokyo City University, Setagaya-ku, Tokyo 158-8557, Japan
}%


\date{\today}

\begin{abstract}
Thermodynamics and hydrodynamics of spontaneous and forced imbibition of liquid into conical capillaries are studied to assess the feasibility of a conical liquid diode. The analytical formulas for the Laplace pressure and the critical Young's contact angle of the capillary for the onset of spontaneous imbibition are derived using the classical capillary model of thermodynamics. The critical contact angle below which the spontaneous imbibition can occur belongs to the hydrophilic region for the capillary with a diverging radius while it belongs to the hydrophobic region for the capillary with a converging radius. Thus, by choosing Young's contact angle between these two critical contact angles, only the spontaneous imbibition toward the converging radius occurs. Therefore, the capillary with a converging radius acts as the forward direction and that with a diverging radius as the reverse direction of diode. Even under the external applied pressure, the free-energy landscape implies that the forced imbibition occurs only to the forward direction by tuning the applied pressure. Furthermore, the scaling rule of the time scale of imbibition is derived by assuming Hagen-Poiseuille steady flow. Again, the time scale of the forward direction is advantageous compared to the reverse direction when the imbibition to both direction is possible. Therefore, our theoretical analysis shows that a conical capillary acts as a liquid diode.
\end{abstract}

\maketitle

%

\section{Introduction}

Capillary imbibition and capillary rise have been studied for centuries and their theoretical foundation was established approximately a century ago~\cite{Bell1906,Lucas1918,Washburn1921,Rideal1922,Bosanquet1923,Landau1987}. Recently, capillary imbibition\cite{Comanns2015,Li2017,Buchberger2018,Urteaga2013,Berli2014,Gorce2016,Singh2020} and droplet transport~\cite{Luo2014,Sen2018,Han2018} into capillaries, grooves and tracs of micro- and nano-scales with geometrical gradients have been attracting much attention, in particular, to investigate the feasibility of the one-way transport devices called the liquid diode.

Conical capillary serves as the simplest model to study the effect of geometrical gradient. It served as the model
of the wetting transition of pores on superhydrophobic substrates~\cite{Nosonovsky2007,Amabili2015,Kaufman2017,Giacomello2019,Iwamatsu2020}, as well as the model of imbibition into
porous substrates~\cite{Staples2002,Young2004,Reyssat2008,Courbin2009}.  It is also pointed out that imbibition into conical capillaries is relevant to the engineering~\cite{Comanns2015,Li2017,Buchberger2018,Urteaga2013,Berli2014,Gorce2016,Singh2020} of the liquid diode: a one-way transport of liquid by the capillary. To develop such a one-way transport micro- and nano-fluidic device, adopting the geometrical~\cite{Urteaga2013,Berli2014,Gorce2016} and chemical gradients~\cite{Singh2020,Panter2020} is the simplest and most efficient approach.

Very recently, for example, Singh et al.~\cite{Singh2020} examined the possibility of the conical capillary as a liquid diode using the dynamical Stokes equation and by assuming the steady capillary flow. Similar strategies are used to study the modification of the scaling rule of imbibition into axisymmetric capillaries~\cite{Reyssat2008,Urteaga2013,Berli2014,Gorce2016} from the original Lucas-Washburn scaling rule of imbibition into cylindrical capillaries~\cite{,Lucas1918,Washburn1921}.

In contrast to those authors who studied the hydrodynamic of imbibition and paid most attention to the time evolution of contact line and the capillary flow, we theoretically consider the imbibition into conical capillaries from the basic thermodynamic perspective.  In fact, several authors~\cite{Tsori2006,Kaufman2017,Panter2020} have already considered imbibition into axisymmetric capillary thermodynamically.  However, those studies paid most attention only to the Laplace pressure at the inlet and the outlet of capillaries.

In this paper, we consider the Laplace pressure and the whole free-energy landscape of imbibition under the applied external pressure from the thermodynamic principle.  We extend our previous study of imbibition into a conical pore of superhydrophobic surface~\cite{Iwamatsu2020} by including a realistic spherical meniscus. We pay special attention to the thermodynamic Laplace pressure because it is the main driving force of capillary flow~\cite{Washburn1921,Landau1987} which can directly answer the question whether or not the  one-way transport is possible~\cite{Tsori2006,Kaufman2017,Panter2020} by the conical capillary. In contrast, the hydrodynamic approach~\cite{Urteaga2013,Berli2014,Gorce2016,Singh2020} can only account for the time scale of the capillary flow when the steady flow can be established.  We also consider the possibility of the liquid diode by the forced imbibition under the action of applied external pressure~\cite{Marmur1988,Dimitrov2008,Schebarchov2011,Gruener2016}.  Finally, we integrate the thermodynamic and the hydrodynamic approaches to understand the asymmetric character of the time scale of the capillary flow of the spontaneous as well as the forced imbibition.  Therefore, the purpose of this paper is to integrate the thermodynamics and the hydrodynamics of capillary imbibition in a conical capillary, and to provide a comprehensive picture that can be useful to design various nano- and micro-fluidic devices including the liquid diode.

\section{\label{sec:sec2}Thermodynamics and Modified Laplace pressure}

\begin{figure}
\includegraphics[width=0.80\linewidth]{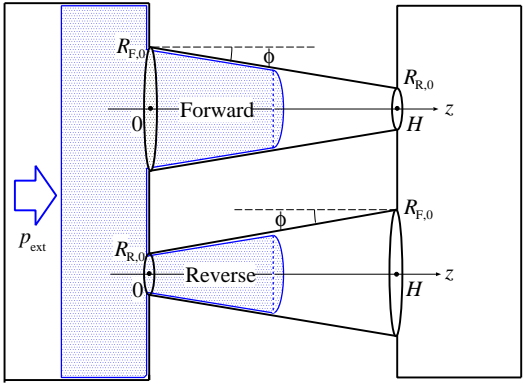}
\caption{\label{fig:1}
Two axial symmetric conical capillaries with (a) a narrowing radius (forward direction) and (b) a widening radius (reverse direction). The opening radii of the conical capillaries are $R_{{\rm F},0}$ and $R_{{\rm R},0}$ for the forward and reverse directions, respectively. The depth of the capillary, rotational axis, and tilt angle of wall are $H$, the $z$-axis, and $\phi$, respectively. The liquid imbibition occurs from left to right. When the imbibition occurs without applied external pressure $p_{\rm ext}$, the spontaneous imbibition occurs.  The forced imbibition could be possible by applying the external pressure.
 }
\end{figure}

The well-known Lucas-Washburn theory~\cite{Lucas1918,Washburn1921}, or sometimes called the Bell-Cameron-Lucas-Washburn~\cite{Bell1906} or Lucas-Washburn-Rideal-Bosanquet~\cite{Rideal1922,Bosanquet1923} theory of capillary imbibition, assumes a fully-developed Hagen-Poiseuille flow described by the parabolic velocity profile in a cylindrical capillary. In this paper, we consider a conical capillary illustrated in Fig.~\ref{fig:1}, and in this section we  consider the thermodynamic of imbibition into such a conical capillary using the classical capillary model.

The surface free energy $F$ comprises the free energy of the free liquid-vapor surface energy $F_{\rm lv}=\gamma_{\rm lv}S_{\rm lv}$ and that of the liquid-solid surface energy of the capillary wall $F_{\rm sl}=\gamma_{\rm lv}\cos\theta_{\rm Y}S_{\rm sl}$ wetted by the liquid. The total surface free energy is given by:
\begin{equation}
F=F_{\rm lv}-F_{\rm sl}=\gamma_{\rm lv}S_{\rm lv}-\gamma_{\rm lv}\cos\theta_{\rm Y}S_{\rm sl},
\label{eq:T1}
\end{equation}
where $\gamma_{\rm lv}$ and $S_{\rm lv}$ represent the liquid-vapor surface tension and surface area, respectively, and $S_{\rm sl}$ is the solid-liquid (wet) surface area. The angle $\theta_{\rm Y}$ is Young's contact angle defined by Young's equation, which is expressed as:
\begin{equation}
\cos\theta_{\rm Y}=\frac{\gamma_{\rm sv}-\gamma_{\rm sl}}{\gamma_{\rm lv}},
\label{eq:T2}
\end{equation}
where $\gamma_{\rm sv}$ and $\gamma_{\rm sl}$ represent the solid-vapor and solid-liquid surface tensions, respectively. This Young's contact angle characterizes the wettability of the capillary wall.   Here, we neglect the effect of gravity since we consider capillaries whose diameters are smaller than the capillary length.  The curvature of the capillary's inner wall is neglected, so that Eq.~(2) represents the force balance in the tangent plane of the conical surface.

We consider an axial symmetric conical capillary with either a narrowing or widening radius (Fig.\ref{fig:1}). We designate the former and the latter as the "Forward" and  "Reverse" directions, respectively, following the electric circuit diode convention, whose radii $R_{\rm F}(z)$ and $R_{\rm R}(z)$ are given by:
\begin{eqnarray}
R_{\rm F}(z) = R_{{\rm F},0}-\tan\phi z,\;\;\;\left(0 \leq z\leq H\right),
\label{eq:T3} \\
R_{\rm R}(z) = R_{{\rm R},0}+\tan\phi z,\;\;\;\left(0 \leq z\leq H\right),
\label{eq:T4}
\end{eqnarray}
where $\phi (0\leq \phi\leq 90^{\circ})$, $R_{{\rm F},0}$ and $R_{{\rm R},0}$, and $H$ represent the tilt angle of the wall, radii at the mouth of the capillary, and length of the capillary (Fig.~\ref{fig:1}).  The "Forward" and "Reverse" designations will be apparent soon.

The capillary parameters in Eqs.~(\ref{eq:T3}) and (\ref{eq:T4}) are related by:
\begin{equation}
R_{{\rm R},0}=R_{{\rm F},0}-\tan\phi H.
\label{eq:T5}
\end{equation}
To study the free-energy landscape along the pathway of imbibition, we have to specify the geometry of conical capillary. We select the tilt angle $\phi$ and aspect ratio $\eta_{\rm F}=H/R_{\rm F,0}$ as the fundamental parameters to specify the geometry. In fact, these two parameters are not independent owing to geometrical constraints $R_{\rm R,0}\ge 0$ and $R_{\rm F,0}>0$ from Eq.~(\ref{eq:T5}), and they satisfy:
\begin{equation}
0<\eta_{\rm F}\le \frac{1}{\tan\phi},
\label{eq:T6}
\end{equation}
where the equality holds when the capillary is a true cone with $R_{\rm R,0}=0$. Figure~2 presents the possible region of the conical capillary in the two-parameter space $\phi$ vs $\eta_{\rm F}$. The aspect ratio $\eta_{\rm F}$ below the line $1/\tan\phi$ is allowed.

\begin{figure}[htbp]
\begin{center}
\includegraphics[width=0.8\linewidth]{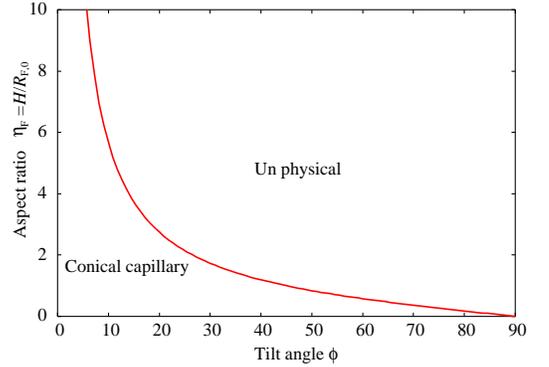}
\end{center}
\caption{
Possible region of conical capillary in the two-parameter space $\phi$ vs $\eta_{\rm F}$. The long capillary with the high aspect ratio $\eta_{\rm F}=H/R_{\rm F,0}$ is possible only when the tilt angle $\phi$ is low.
 } 
\label{fig:2}
\end{figure}

We consider the surface free energy $F$ when the position of the liquid-vapor meniscus in the pore is $z$ (Fig.~\ref{fig:1}). Although the detailed formulation for a simpler model has already been published~\cite{Iwamatsu2020}, we summarize the main result briefly. Because we are more interested in the asymmetry of the {\it same shape}, we will adopt suffix $i$ for $i=$"F" and "R'' for the forward and reverse directions, respectively. Hence, we will use suffix $i$ to represent the forward and reverse directions collectively.

\begin{figure}[htbp]
\begin{center}
\includegraphics[width=0.9\linewidth]{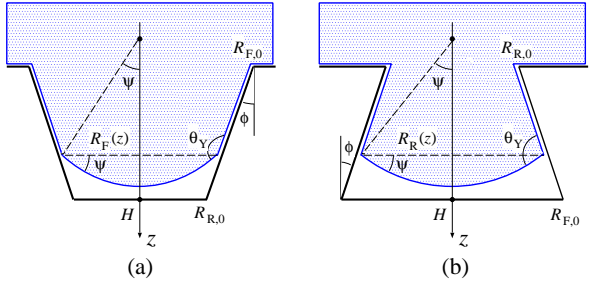}
\end{center}
\caption{
Spherical meniscus of liquid in the conical capillary of (a) forward and (b) reverse directions.  The opening angle $\psi$ is related to the tilt angle $\phi$ and Young's contact angle $\theta_{\rm Y}$.  Here, we show the convex meniscus with the contact angle $\theta_{\rm Y}$ larger than $90^{\circ}$.  In fact, the meniscus must be concave to make the modified Laplace pressure positive and the spontaneous imbibition possible. 
} 
\label{fig:3}
\end{figure}

The solid-liquid surface free energy is given by~\cite{Iwamatsu2020}:
\begin{eqnarray}
F_{i,{\rm sl}} &=& 2\pi \gamma_{\rm lv} \cos \theta_{\rm Y}  \int_{0}^{z} R_{i}(z')\sqrt{1+\left(\frac{dR_{i}}{dz'}\right)^{2}}dz'
\nonumber \\
&=& 2\pi\gamma_{\rm lv}\frac{\cos\theta_{\rm Y}}{\cos\phi}  \int_{0}^{z} R_{i}(z') dz',
\label{eq:T7}
\end{eqnarray}
for $i=$"F"and "R.''  The liquid-vapor surface free energy is given by:
\begin{equation}
F_{i,{\rm lv}} = 2\pi \gamma_{\rm lv}R_{i}(z)^2\frac{1-\cos\psi}{\sin^{2}\psi},
\label{eq:T8}
\end{equation}
where the opening angles $\psi$ defined in Fig.~\ref{fig:3} are given by:
\begin{equation}
\psi=\theta_{\rm Y}-\phi-90^{\circ}
\label{eq:T9}
\end{equation}
for the forward direction $i=$"F," and
\begin{equation}
\psi=\theta_{\rm Y}+\phi-90^{\circ}
\label{eq:T10}
\end{equation}
for the reverse direction $i=$"R."  In contrast to the previous study~\cite{Iwamatsu2020}, we assume the spherical liquid-vapor interface.

The thermodynamic liquid pressure $p_{i,{\rm L}}(z)$ is defined by
\begin{equation}
p_{i,{\rm L}}(z)=-\frac{\partial F_{i}}{\partial V_{i}}=-\frac{1}{dV_{i}/dz}\frac{\partial F_{i}(z)}{\partial z},
\label{eq:T11}
\end{equation}
where
\begin{equation}
V_{i}(z)=\pi\int_{0}^{z}R_{i}(z')^{2}dz'+\frac{\pi}{3}R_{i}(z)^{3}\nu_{i}\left(\psi\right)
\label{eq:T12}
\end{equation}
is the total volume of the liquid inside the capillary, and
\begin{equation}
\nu_{i}\left(\psi\right)=\frac{\left(1-\cos\psi\right)^{2}\left(2+\cos\psi\right)}{\sin^{3}\psi}
\label{eq:T13}
\end{equation}
is the small volume correction from the spherical liquid-vapor interface, which is explicitly given by
\begin{equation}
\nu_{i}\left(\theta_{\rm Y},\phi\right) =-\frac{\left(1-\cos\left(\theta_{\rm Y}\mp\phi\right)\right)^{2}\left(2+\sin\left(\theta_{\rm Y}\mp\phi\right)\right)}{\cos^{3}\left(\theta_{\rm Y}\mp\phi\right)},
\label{eq:T14} 
\end{equation}
because the opening angle $\psi$ is given either by Eqs.~(\ref{eq:T9}) (forward) or (\ref{eq:T10}) (reverse). The upper $-$ in $\mp$ is for the forward direction $i=$F and the lower $+$ is for the reverse direction $i=$R.  To reduce the number of equations and to save the space, hereafter we use the upper sign $-$ or $+$ in $\mp$ or $\pm$ to indicate the forward and the lower sign $+$ or $-$ to indicate the reverse direction.

Then,
\begin{equation}
\frac{dV_{i}}{dz}=\pi R_{i}(z)^{2}\left(1+\frac{dR_{i}}{dz}\nu_{i}\left(\theta_{\rm Y},\phi\right)\right),
\label{eq:T15}
\end{equation}
and, Eq.~(\ref{eq:T11}) can be written in the form of the modified Laplace pressure as
\begin{equation}
p_{i,{\rm L}}(z)=\frac{2\gamma_{\rm lv}\Pi_{i}\left(\theta_{\rm Y},\phi\right)}{R_{i,{\rm eff}}(z)},
\label{eq:T16}
\end{equation}
where the scaled non-dimensional pressure $\Pi_{i}$ is given by:
\begin{equation}
\Pi_{i}\left(\theta_{\rm Y},\phi\right) = \frac{2\tan\phi}{1+\sin\left(\theta_{\rm Y}\mp\phi\right)}+\frac{\cos\theta_{\rm Y}}{\cos\phi},
\label{eq:T17}
\end{equation}
for the forward $i=$F and reverse $i=$R directions, respectively, where we have eliminated $\psi$ using Eqs.~(\ref{eq:T9}) and (\ref{eq:T10}).  Note that the scaled pressure $\Pi_{\rm R}$ is obtained simply by replacing the sign of $\phi$ ($\phi\rightarrow -\phi$) of $\Pi_{\rm F}$, and vice versa.  The volume-corrected effective radius $R_{i,{\rm eff}}$ defined by Eq.~(\ref{eq:T16}) is written as
\begin{equation}
R_{i, {\rm eff}}(z) = R_{i}(z)\left(1\mp\tan\phi\nu_{i}\left(\theta_{i},\phi\right)\right),
\label{eq:T18}
\end{equation}
for the forward and the reverse directions from Eq.~(\ref{eq:T15}).  A formula similar to Eq.~(\ref{eq:T17}) has been derived as the local force balance equation at the opening of the capillary in the context of superhydrophobic surfaces~\cite{Kaufman2017}.

This liquid pressure $p_{i,{\rm L}}(z)$ can be positive or negative depending on the sign of the non-dimensional pressures in Eq.~(\ref{eq:T17}).  Spontaneous imbibition is possible only when this pressure $p_{i,{\rm L}}(z)$ or $\Pi_{i}\left(\theta_{\rm Y},\phi\right)$ is positive.  Otherwise, applied pressure is necessary to force the liquid to penetrate into capillaries.

The absolute magnitude of this pressure $\left| p_{i,{\rm L}} \right|$ depends on the meniscus position $z$ through the effective radius $R_{i, {\rm eff}}(z)$ or the capillary radius $R_{i}(z)$ in Eq.~(\ref{eq:T18}).  Hence, the absolute magnitude of this driving pressure will be maximum when the radius become smallest when $R_{i}(z)=R_{\rm R,0}$, which occurs at the capillary outlet at $z=H$ for the forward direction and at the capillary inlet at $z=0$ for the reverse direction.

When the liquid pressure is negative, the external pressure to overcome this negative pressure is necessary to force the imbibition.  Then, the highest external pressure is necessary at the outlet of the capillary for the forward direction and at the inlet for the reverse direction.

The liquid pressure in Eq.~(\ref{eq:T16}) reduces to the standard Laplace pressure,
\begin{equation}
p_{i,{\rm L}}=\frac{2\gamma_{\rm lv}\cos\theta_{\rm Y}}{R_{i}}
\label{eq:T19}
\end{equation}
in straight cylinders with $\phi=0^{\circ}$ and $R_{i}={\rm constant}$.  The spontaneous imbibition is possible only when $p_{i,{\rm L}}\ge 0$ and the capillary must be hydrophilic with $\theta_{\rm Y}\leq 90^{\circ}$ in straight cylinders.  In the conical capillary, however, the sign of the liquid pressure in Eq.~(\ref{eq:T16}) depends not only on the wettability $\theta_{\rm Y}$ but also on the tilt angle $\phi$ through the scaled non-dimensional pressures $\Pi_{\rm F}\left(\theta_{\rm Y},\phi\right)$ and $\Pi_{\rm R}\left(\theta_{\rm Y},\phi\right)$ in Eq.~(\ref{eq:T17}).

\begin{figure}[htbp]
\begin{center}
\includegraphics[width=0.8\linewidth]{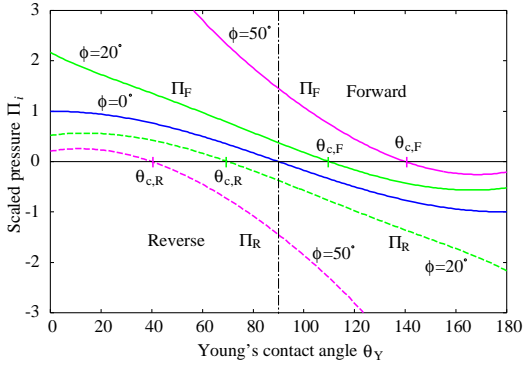}
\end{center}
\caption{
Scaled pressures $\Pi_{\rm F}\left(\theta_{\rm Y},\phi\right)$ (solid lines) and $\Pi_{\rm R}\left(\theta_{\rm Y},\phi\right)$ (broken lines) as a function of Young's contact angle $\theta_{\rm Y}$ for a low tilt angle $\phi=20^{\circ}$ and a high tilt angle $\phi=50^{\circ}$, respectively. For comparison, the scaled pressure for the straight cylinder with $\phi=0^{\circ}$ is depicted.  The pressure is positive when Young's angle $\theta_{\rm Y}$ is smaller than the critical angles $\theta_{\rm c,F}$ and $\theta_{\rm c,R}$. Spontaneous imbibition is possible when this pressure is positive. Accordingly, spontaneous imbibition is possible when $\theta_{\rm Y}<\theta_{\rm c,R}$ for the reverse direction and $\theta_{\rm Y}<\theta_{\rm c,F}$ for the forward direction.  Note pronounced asymmetry between the forward direction F and the reverse direction R, which is enhanced at a higher tilt angle $\phi$.
} 
\label{fig:4}
\end{figure}

Figure \ref{fig:4} presents the scaled non-dimensional pressures $\Pi_{\rm F}\left(\theta_{\rm Y},\phi\right)$ and $\Pi_{\rm R}\left(\theta_{\rm Y},\phi\right)$ as a function of Young's contact angle $\theta_{\rm Y}$ for a low tilt angle $\phi=20^{\circ}$ and a high tilt angle $\phi=50^{\circ}$, respectively. Spontaneous imbibition is possible when this pressure is positive.  Apparently, spontaneous imbibition for the forward direction can occur even if the substrate is hydrophobic ($\theta_{\rm Y}>90^{\circ}$) while that for the reverse direction can occur only if the substrate is hydrophilic  ($\theta_{\rm Y}<90^{\circ}$).  A formula similar to Eq.~(\ref{eq:T17}) and a diagram similar to Fig.~\ref{fig:4} for the local force balance at the opening of capillary in the context of superhydrophobic surfaces have already been presented~\cite{Kaufman2017}.  Here, however, we consider the liquid pressure to account for the possibility of one-way transport in conical capillaries.

When this pressure becomes negative ($p_{i,{\rm L}}<0$ or $\Pi_{i}<0$), capillary imbibition will be prohibited because no driving pressure exists. This critical contact angle for the forward direction $i=$F is determined from the condition $p_{i,{\rm L}}=0$ or $\Pi_{i}=0$ (Fig.~\ref{fig:4}), which leads to
\begin{equation}
2\sin\phi+\left(1+\sin\left(\theta_{\rm c,F}-\phi\right)\right)\cos\theta_{\rm c,F}=0
\label{eq:T20}
\end{equation}
from Eq.~(\ref{eq:T17}), whose solution is $\theta_{\rm c,F}-\phi=90^{\circ}$ or
\begin{equation}
\theta_{\rm c,F}=90^{\circ}+\phi.
\label{eq:T21}
\end{equation}
Likewise, for the reverse direction $i=$R, this critical contact angle is determined from
\begin{equation}
-2\sin\phi+\left(1+\sin\left(\theta_{\rm c,R}+\phi\right)\right)\cos\theta_{\rm c,R}=0
\label{eq:T22}
\end{equation}
from Eq.~(\ref{eq:T17}), whose solution is $\theta_{\rm c,R}+\phi=90^{\circ}$ or
\begin{equation}
\theta_{\rm c,R}=90^{\circ}-\phi.
\label{eq:T23}
\end{equation}

At these critical contact angles $\theta_{\rm c,F}$ and $\theta_{\rm c,R}$, the opening angle $\psi$ vanishes ($\psi=0^{\circ}$) from Eqs.~(\ref{eq:T9}) and (\ref{eq:T10}). The meniscus becomes flat and the free-energy cost to increase or decrease the liquid volume vanishes because $\partial F_{i}/\partial V_{i}=0$. Then, the liquid-vapor interface will be delocalized and the liquid fills the capillary by the mechanism known as the filling transition of the wedge and cone~\cite{Hauge1992,Reijmer1999,Malijevsky2015}, even though the driving pressure to induce the liquid flow is absent ($p_{i,{\rm L}}=-\partial F_{i}/\partial V_{i}=0$).  Apparently, the capillary must be hydrophilic ($\theta_{\rm Y}<90^{\circ}$) for the reverse direction, but it can be hydrophobic ($\theta_{\rm Y}>90^{\circ}$) for the forward direction, which is intuitively apparent from Figs.~\ref{fig:3}(a) and \ref{fig:3}(b) when the meniscus is flat.

Our results in Eqs.~(\ref{eq:T21}) and (\ref{eq:T23}) are exactly the same as those derived from the routinely used Laplace pressure derived from the mechanical consideration:
\begin{equation}
p_{i,{\rm L}}(z)=\frac{2\gamma_{\rm lv}\cos\left(\theta_{\rm Y}\mp\phi\right)}{R_{i}(z)}.
\label{eq:T24}
\end{equation}
Our expression in Eq.~(\ref{eq:T16}), however, is based on the thermodynamics, and we can consider the free-energy landscape of imbibition, which is the subject of Sec.~\ref{sec:sec3}.

\begin{figure}[htbp]
\begin{center}
\includegraphics[width=0.8\linewidth]{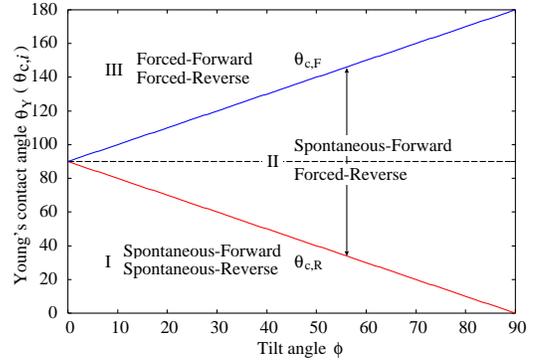}
\end{center}
\caption{
Critical Young's contact angles $\theta_{\rm c, F}$ and $\theta_{\rm c, R}$ as functions of the tilt angles $\phi$ given by Eqs.~(\ref{eq:T21}) and (\ref{eq:T23}). The former belongs to the hydrophobic region and the latter belongs to the hydrophilic region. The spontaneous imbibition is possible only below these lines, otherwise the applied pressure is necessary to force the imbibition.  The $\left(\theta_{\rm Y},\phi\right)$ space is divided into three regions I, II and III by these two lines.  Only the spontaneous imbibition to the forward direction (liquid diode) is possible in the region II.  In region I, the spontaneous imbibition to both directions are possible, while only the forced imbibition to both direction is possible in region III.
 } 
\label{fig:5}
\end{figure}

Figure~\ref{fig:5} represents the critical contact angles $\theta_{\rm c, F}$ and $\theta_{\rm c, R}$ as a function of the tilt angle $\phi$. The critical contact angle for the forward direction belongs to the hydrophobic region $\theta_{\rm c, F}>90^{\circ}$, whereas that for the reverse direction belongs to the hydrophilic region $\theta_{\rm c, R}<90^{\circ}$ (Fig.~\ref{fig:4}).  In particular, when $\phi=0^{\circ}$ (a straight cylinder), $\theta_{\rm c, F}=\theta_{\rm c, R}=90^{\circ}$ such that imbibition can only occur in the hydrophilic capillary.

The $\left(\theta_{\rm Y},\phi\right)$ space in Fig.~\ref{fig:5} is divided into three regions I, II and III by two lines defined by Eqs.~(\ref{eq:T21}) and (\ref{eq:T23}).  In region I when Young's contact angle $\theta_{\rm Y}$ is smaller than the critical angle $\theta_{\rm c,R}$, it is also smaller than $\theta_{\rm c,F}$ (Fig.~\ref{fig:3}).  Therefore, the spontaneous imbibition for both directions is permitted because $\theta_{\rm Y}<\theta_{\rm c,R}<\theta_{\rm c,F}$.

In region II when $\theta_{\rm c,R}<\theta_{\rm Y}<\theta_{\rm c,F}$, the spontaneous imbibition only for the forward direction is possible. Therefore, in this region, the conical capillary functions as a liquid diode.  In addition, Young's contact angle $\theta_{\rm Y}$ can be larger than $90^{\circ}$. To expand the range of the diode, a larger tilt angle $\phi$ is advantageous from Fig.~\ref{fig:5}. However, this indicates a short capillary of low aspect ratio from Fig.~\ref{fig:2}.

In region III when Young's contact angle $\theta_{\rm Y}$ is larger than the critical angle for the forward direction, it is also larger than the critical angle for the reverse direction  ($\theta_{\rm c,R}<\theta_{\rm c,F}<\theta_{\rm Y}$).  Consequently, the spontaneous imbibition is prohibited for both directions.  Only forced imbibition, which is realized by applying the external pressure (Fig.~\ref{fig:1}) to the liquid, is possible. Further investigation of the free-energy landscape~\cite{Iwamatsu2020} will be necessary to judge the possibility of the forced imbibition process, which is presented in Sec.~\ref{sec:sec3}.

So far, we have been considering the geometrical gradient where the geometry (radius) of the capillary changes gradually.
It is straightforward to extend our analysis to the chemical gradient~\cite{Singh2020,Panter2020} where the surface wettability of the capillary changes gradually, which can be accounted for by including the site-dependent contact angle $\theta_{\rm Y}(z)$ in Eq.~(\ref{eq:T7}).  It is straightforward to show that the liquid pressure, which corresponds to Eq.~(\ref{eq:T16}) is given by Eq.~(\ref{eq:T19}) with local Young's contact angle $\theta_{\rm Y}(z)$ when only the chemical gradient exists.  There is no distinction between the forward and the reverse direction.  Therefore, the geometrical gradient is always necessary to realize a device of one-way spontaneous imbibition.

In this section, we used the thermodynamic approach and derived an analytical formula for the modified Laplace pressure from which we determined a criterion for the diodic character (one-way transport) in conical capillaries.  In fact, many researchers~\cite{Urteaga2013,Berli2014,Gorce2016,Singh2020} used the hydrodynamic model~\cite{Reyssat2008}, and compared the time scale of flow in the forward and the reverse direction.  A large difference in time scale between them was considered as the diodic character of conical capillaries.  However, such a hydrodynamic study can be meaningful only when the spontaneous imbibition of  both the forward and reverse directions is possible which is represented by region I of  Fig.~{\ref{fig:5}}.  In contrast, our thermodynamic study directly showed that the one-way spontaneous imbibition is possible only in region II of Fig.~\ref{fig:5}. Even when the spontaneous imbibition is prohibited, it is possible to induce the capillary flow by applied external pressure.  The question whether or not the one-way forced imbibition is possible in region III of Fig.~\ref{fig:5} will be answered in Sec.~\ref{sec:sec3}.

\section{\label{sec:sec3}Thermodynamics of forced imbibition}

In region III of Fig.~\ref{fig:5} when the driving pressure $p_{i,{\rm L}}(z)$ or the scaled pressure $\Pi_{i}\left(\theta_{\rm Y},\phi\right)$ is negative, the spontaneous imbibition is prohibited. Applied external pressure $p_{\rm ext}$ is necessary to push the liquid into the capillary. Before considering the forced imbibition by external applied pressure, we consider the free-energy landscape $F_{i}$ in Eq.~(\ref{eq:T1}) of the spontaneous imbibition in regions I and II in Fig.~\ref{fig:5} when $p_{\rm ext}=0$.

Because the integrals in Eqs.~(\ref{eq:T7}) and (\ref{eq:T12}) can be easily evaluated, the excess free energy
\begin{equation}
\Delta F_{i}(z)=F_{i}(z)-F_{i}(0)
\label{eq:T25}
\end{equation}
becomes a quadratic function of the meniscus position $z$ after mathematical manipulation, and it is given by
\begin{equation}
\Delta F_{i}(z)=\pi R_{i,0}^{2} \frac{2\gamma_{\rm lv}\Pi_{i}\left(\theta_{\rm Y},\phi\right)}{R_{i,0}}\left(-z\pm\frac{1}{2}\frac{\tan\phi}{R_{i,0}}z^{2}\right),
\label{eq:T26}
\end{equation}
where the upper $+$ in $\pm$ is for the forward direction $i=$F and the lower $-$ is for the reverse direction $i=$R.  The free energy $\Delta F_{\rm R}$ is obtained from $\Delta F_{\rm F}$ by replacing $\phi\rightarrow -\phi$.

By introducing the characteristic driving pressure similar to Eq.~(\ref{eq:T16}) defined by
\begin{equation}
\bar{p}_{i,0}=\frac{2\gamma_{\rm lv}\Pi_{i}\left(\theta_{\rm Y},\phi\right)}{R_{i,0}},
\label{eq:T27}
\end{equation}
which does not depend on the meniscus position $z$, Eq.~(\ref{eq:T26}) is written in more compact and transparent form:
\begin{equation}
\Delta F_{i}(z)=\pi R_{i,0}^{2} \bar{p}_{i,0} \left(-z\pm\frac{1}{2}\frac{\tan\phi}{R_{i,0}}z^{2}\right).
\label{eq:T28}
\end{equation}
Young's contact angle $\theta_{\rm Y}$ is implicitly included in the characteristic driving pressure $\bar{p}_{i,0}$ in Eq.~(\ref{eq:T27}), whose sign is always the same as that of the modified Laplace pressure $p_{i,{\rm L}}(z)$ in (\ref{eq:T16}). The free energy always starts to decrease at $z=0$ when the spontaneous imbibition occurs when $p_{i,{\rm L}}(z)>0$ or $\bar{p}_{i,0}>0$ from Eq.~(\ref{eq:T28}).

At the critical contact angles $\theta_{\rm c,F}$ and $\theta_{\rm c,R}$ defined by Eqs.~(\ref{eq:T21}) and (\ref{eq:T23}), the characteristic driving pressure vanish ($\bar{p}_{i,0}=0$) because $\Pi_{i}\left(\theta_{{\rm c},i},\phi\right)=0$. The free energy neither increases nor decreases [$\Delta F_{i}(z)=0$]. The cost of the surface free energy to move a flat liquid-vapor interface vanishes. Then, the liquid-vapor interface will be delocalized and the liquid fills the capillary not by the mechanism of imbibition but by the mechanism known as the filling transition of the wedge and cone~\cite{Hauge1992,Reijmer1999,Malijevsky2015}.

The extremum of the free energy for the reverse direction at $z_{\rm ex}=-2R_{{\rm R},0}/\tan\phi$ and that for the forward direction at $z_{\rm ex}=2R_{{\rm F},0}/\tan\phi$ from Eq.~(\ref{eq:T28}) belongs to the unphysical domains $z_{\rm ex}<0$ or $z_{\rm ex}>H$. Therefore, the free energy is a monotonically decreasing function of $z$ in the domain $0\leq z\leq H$ when the spontaneous imbibition occurs ($p_{i,{\rm L}}(z)>0, \bar{p}_{i,0}>0$) in regions I and II of Fig.~\ref{fig:5}.

When the driving pressure is negative, $p_{i,{\rm L}}(z)<0$ or $\Pi_{i}\left(\theta_{\rm Y},\phi\right)<0$, the spontaneous imbibition is prohibited because the free energy $\Delta F_{i}(z)$ is an increasing function of the meniscus position $z$ as $\bar{p}_{i,0}<0$. It is necessary to apply external pressure $p_{\rm ext}>0$ to overcome this negative liquid pressure to force the imbibition. The thermodynamics of this process is described by the free energy called grand potential given by
\begin{equation}
G_{i}=F_{i}-p_{\rm ext}V_{i},
\label{eq:T29}
\end{equation}
instead of Helmholtz free energy in Eq.~(\ref{eq:T1}), where $V_{i}$ is the liquid volume inside the capillary given by Eq.~(\ref{eq:T12}). Then, the driving pressure of capillary flow is given by
\begin{equation}
p_{i}(z)=-\frac{\partial G_{i}}{\partial V_{i}}=p_{\rm ext}+p_{i,{\rm L}}(z),
\label{eq:T30}
\end{equation}
where $p_{i,{\rm L}}(z)$ is the modified Laplace pressure given by Eq.~(\ref{eq:T16}). Therefore, the driving pressure $p_{i}(z)$ can be positive even when the modified Laplace pressure is negative [$p_{i,{\rm L}}(z)<0$] if $p_{\rm ext}>-p_{i,{\rm L}}(z)$.

Because the absolute magnitude of the modified Laplace pressure $p_{i, {\rm L}}(z)$ becomes maximum when the capillary radius is smallest ($R_{i}(z)=R_{\rm R,0}$), the driving pressure can be always positive during the imbibition of capillary when
\begin{equation}
p_{\rm ext} > -p_{\rm F,L}(z=H) ,
\label{eq:T31}
\end{equation}
for the forward direction, and
\begin{equation}
p_{\rm ext} > -p_{\rm R,L}(z=0) 
\label{eq:T32}
\end{equation}
for the reverse direction. Therefore, the minimum external pressure to force the imbibition is the modified Laplace pressure at the outlet of the capillary for the forward direction and that at the inlet of the capillary for the reverse direction.

To understand the detailed process of imbibition, it is useful to know the excess free-energy landscape $G_{i}(z)$.  By integrating Eq.~(\ref{eq:T12}), we obtain
\begin{eqnarray}
p_{\rm ext}V_{i} &=& p_{\rm ext}\frac{\pi}{3}\nu_{i}\left(\theta_{\rm Y}\mp\phi\right)R_{i,0}^{3}
\nonumber \\
&+& \pi R_{i,0}^{2}p_{\rm ext}\left(1-\nu_{i}\left(\theta_{\rm Y}\mp\phi\right)\tan\phi\right)
\nonumber \\
&\times& \left(z\mp\frac{\tan\phi}{R_{i,0}}z^{2}+\frac{1}{3}\left(\frac{\tan\phi}{R_{i,0}}\right)^{2}z^{3}\right)
\label{eq:T33}
\end{eqnarray}
for the forward (upper $-$) and the reverse (lower $+$) directions. By combining Eqs.(\ref{eq:T28}) and (\ref{eq:T33}) we obtain the excess free energy $\Delta G_{i}$,
\begin{eqnarray}
\Delta G_{i}(z) &=& \pi R_{i,0}^{2} \left(-\left(\bar{p}_{i,{\rm ext}} + \bar{p}_{i,0}\right)z \right.
\nonumber \\
&\pm& \left.\left(\bar{p}_{i,{\rm ext}}+\frac{\bar{p}_{i,0}}{2}\right)\frac{\tan\phi}{R_{i,0}}z^{2} \right.
\nonumber \\
&-& \left. \frac{1}{3}\bar{p}_{\rm ext}\left(\frac{\tan\phi}{R_{i,0}}\right)^{2}z^{3}\right),
\label{eq:T34}
\end{eqnarray}
for the forward (upper $+$) and the reverse (lower $-$) directions, where
\begin{equation}
\bar{p}_{i,{\rm ext}}=\left(1\mp\nu_{i}\left(\theta_{\rm Y}\mp\phi\right))\tan\phi\right)p_{\rm ext}
\label{eq:T35}
\end{equation}
is the external pressure modified by the volume correction $\nu_{i}$ and the characteristic pressure $\bar{p}_{i,0}$ is defined by Eq.~(\ref{eq:T27}). This free energy in Eq.~(\ref{eq:T34}) given by a cubic polynomial in $\tilde{z}$ has exactly the same form as  the free energy derived in our previous study~\cite{Iwamatsu2020} by assuming a flat liquid-vapor meniscus.

The forced imbibition starts at the capillary inlet at $z=0$ when 
\begin{equation}
\bar{p}_{i,{\rm ext}}>-\bar{p}_{i,0}=-\frac{2\gamma_{\rm lv}\Pi_{i}\left(\theta_{\rm Y},\phi\right)}{R_{i,0}}
\label{eq:T36}
\end{equation}
from the linear term of Eq.~(\ref{eq:T34}). Note that $\bar{p}_{i,{\rm ext}}>0$ and $\bar{p}_{i,0}<0$. This condition (with $i=$R) reduces to Eq.~(\ref{eq:T32}), while it does not reduce to Eq.~(\ref{eq:T31}). In fact, Eq.~(\ref{eq:T31}) is written as
\begin{equation}
\bar{p}_{\rm F, ext}>-\frac{2\gamma_{\rm lv}\Pi_{\rm F}\left(\theta_{\rm Y},\phi\right)}{R_{\rm R,0}}=-\frac{R_{\rm F,0}}{R_{\rm R,0}}\bar{p}_{{\rm F},0}>-\bar{p}_{{\rm F},0},
\label{eq:T37}
\end{equation}
which is the driving pressure not at the inlet but at the outlet of the capillary.

Therefore, once the imbibition starts at the inlet of the capillary, the liquid will continue to invade the capillary until the meniscus reaches the outlet for the reverse direction. However, even if the external pressure $\bar{p}_{\rm ext}$ satisfies the inequality (\ref{eq:T36}) and the liquid starts to invade at $z=0$ for the forward direction, the driving pressure would vanish or become negative in the middle of the capillary because the external pressure $\bar{p}_{\rm ext}$ does not satisfy the inequality (\ref{eq:T37}) when
\begin{equation}
-\frac{R_{\rm F,0}}{R_{\rm R,0}}\bar{p}_{{\rm F},0}>\bar{p}_{\rm ext}>-\bar{p}_{{\rm F},0}.
\label{eq:T38}
\end{equation}
Strong asymmetry between the imbibition for the forward and the reverse directions exists in the forced imbibition as well.

To understand the detailed process of imbibition, it is useful to know the excess free-energy landscape $\Delta G_{i}\left(z\right)$.  To this end, we define the non-dimensional quantities
\begin{eqnarray}
\tilde{z} &=& \frac{z}{H},
\label{eq:T39} \\
\tilde{p} &=& \frac{\bar{p}_{\rm ext}}{\left|\bar{p}_{\rm F,0}\right|},
\label{eq:T40} \\
\alpha_{\rm F} &=& \eta_{\rm F}\tan\phi = \frac{H\tan\phi}{R_{\rm F,0}},
\label{eq:T41} \\
\Delta \tilde{{\rm g}}_{\rm F} &=& \frac{\Delta G_{\rm F}}{\pi R_{\rm F,0}^{2}\left|\bar{p}_{\rm F,0}\right| H}
\label{eq:T42}
\end{eqnarray}
for the forward direction so that we can reduce the number of parameters and compare the result directly to those derived in our previous paper.~\cite{Iwamatsu2020}

Then, the free-energy landscape of forced imbibition in Eq.~(\ref{eq:T34}) when $\bar{p}_{\rm F,0}<0$ is written as
\begin{equation}
\Delta \tilde{{\rm g}}_{\rm F}\left(\tilde{z}\right)=\left(1-\tilde{p}\right)\tilde{z}+\left(\tilde{p}-\frac{1}{2}\right)\alpha_{\rm F}\tilde{z}^{2}-\frac{1}{3}\tilde{p}\alpha_{\rm F}^{2}\tilde{z}^{3},
\label{eq:T43}
\end{equation}
which has exactly the same form as that derived in our previous work for the simplified model with a flat meniscus~\cite{Iwamatsu2020}. Therefore, the effect of the curvature of meniscus does not change the character of the free-energy landscape.  The free energy of spontaneous imbibition in Eq.~(\ref{eq:T26}) with $\bar{p}_{\rm F,0}>0$ is simply given by changing the sign as $\Delta \tilde{{\rm g}}_{\rm F}\rightarrow -\Delta \tilde{{\rm g}}_{\rm F}$ from Eq.~(\ref{eq:T42}) and setting $\tilde{p}=0$.  This free energy in Eq.~(\ref{eq:T43}) is a cubic polynomial of $\tilde{z}$ ~\cite{Iwamatsu2020}.

The imbibition starts at $\tilde{z}=0$ when $\tilde{p}>1$, which corresponds to inequality (\ref{eq:T36}). The free-energy landscape has two extremums at $\alpha_{\rm F}\tilde{z}=1$ and $\alpha_{\rm F}\tilde{z}=\left(\tilde{p}-1\right)/\tilde{p}$ from $d\Delta \tilde{{\rm g}}_{\rm F}/d\tilde{z}=0$. The former belongs to the unphysical domain $\tilde{z}=1/\alpha_{\rm F}>1$ from Eqs.~(\ref{eq:T6}) and (\ref{eq:T41}). The latter corresponds to the free-energy minimum~\cite{Iwamatsu2020} at
\begin{equation}
\tilde{z}_{\rm min}=\frac{\tilde{p}-1}{\alpha_{\rm F}\tilde{p}},
\label{eq:T44}
\end{equation}
where the liquid pressure $p_{i}(z)$ in Eq.~(\ref{eq:T30}) changes sign from positive to negative.  The liquid flow stops at the middle of the conical capillary $z_{\rm min}$.  The steady capillary flow for the forward direction will not be established even under the applied external pressure as long as $\tilde{z}_{\rm min}<1$ or $z_{\rm min}<H$ from Eq.~(\ref{eq:T39}).

\begin{figure}[htbp]
\begin{center}
\includegraphics[width=0.8\linewidth]{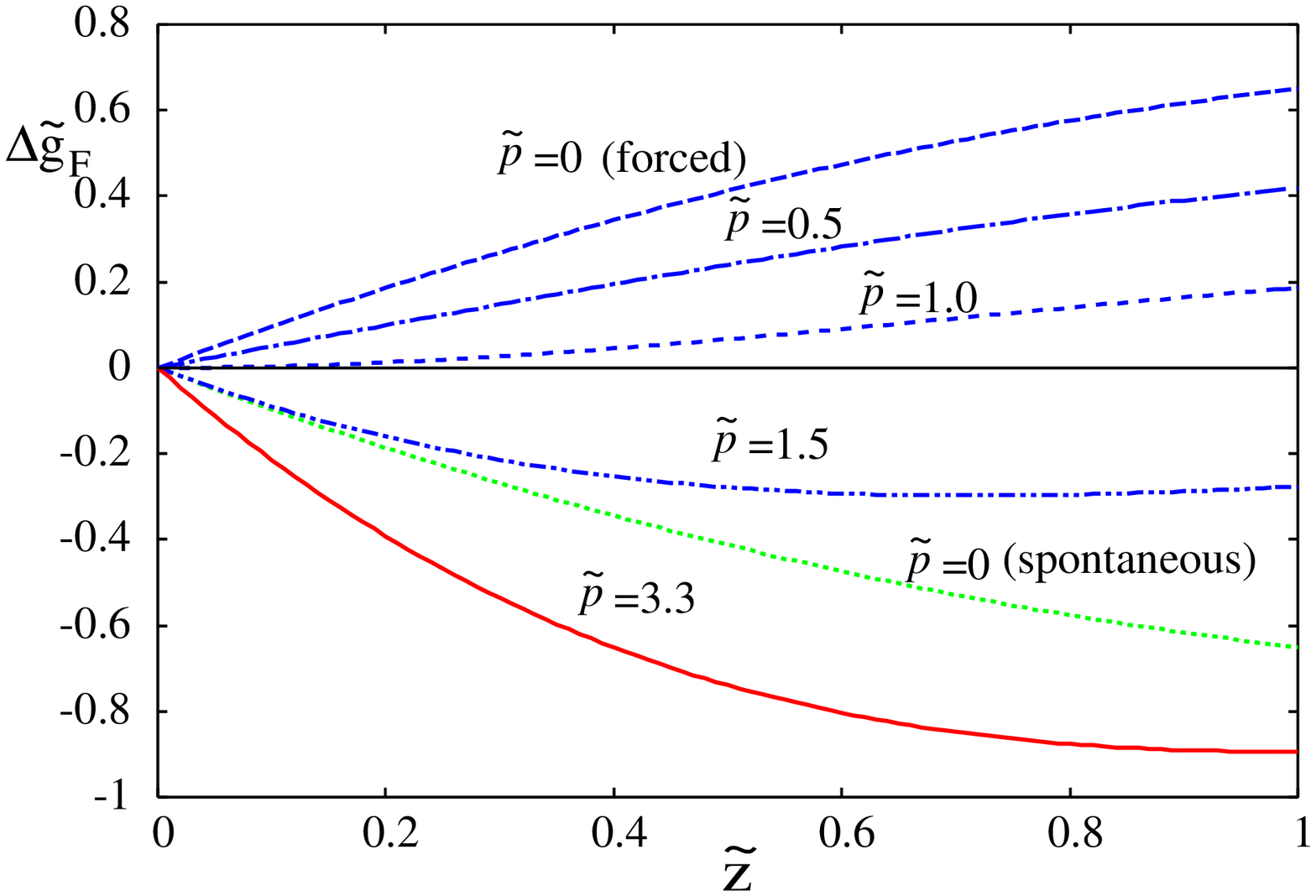}
\end{center}
\caption{
The free-energy landscape $\Delta \tilde{\rm g}_{\rm F}\left(\tilde{z}\right)$ for the forward direction in Eq.~(\ref{eq:T43}). The free-energy landscapes of the forced and the spontaneous imbibition with no applied pressure $\tilde{p}=0$ are also shown. The geometrical parameter is $\alpha_{\rm F}=0.7$ so that $\alpha_{\rm R}=2.33$. The free-energy landscape for the forward direction exhibits minimum when $\tilde{p}\ge 1$.  The forced imbibition starts at the inlet of the capillary when $\tilde{p}>1$  from Eq.~(\ref{eq:T36}) for both the forward and the reverse direction.  However, the forced imbibition for the forward direction throughout the capillary is established only when $\tilde{p}>3.3$ given by Eq.~(45).
}
\label{fig:6}
\end{figure}

Figure~\ref{fig:6} presents the free-energy landscape $\Delta \tilde{{\rm g}}_{\rm F}\left(\tilde{z}\right)$ for various reduced pressures $\tilde{p}$. We also show the free-energy landscape when $\tilde{p}=0$ and $\bar{p}_{\rm F,0}<0$ (forced imbibition) as well as $\bar{p}_{\rm F,0}>0$ (spontaneous imbibition), whose free energy (in original unit) is given by Eq.~(\ref{eq:T28}). The free-energy landscape shows minimum at $\tilde{z}_{\rm min}=0$ when $\tilde{p}=1$ ($\bar{p}_{\rm ext}=\left|\bar{p}_{\rm F,0}\right|$). Then the forced imbibition starts at the inlet of the capillary at $\tilde{z}=0$. However, the free energy increases as the meniscus position $\tilde{z}$ moves from the inlet.  The liquid pressure $p_{\rm F}(z)$ in Eq.~(\ref{eq:T30}) becomes negative and the meniscus cannot move from $\tilde{z}=0$. By further increasing the external pressure (e.g. $\tilde{p}=1.5$ in Fig.~\ref{fig:6}), the minimum position $\tilde{z}_{\rm min}$ moves into the capillary. However, the driving capillary pressure $p_{\rm F}(z)$ in Eq.~(\ref{eq:T30}) turns from positive to negative at $\tilde{z}_{\rm min}$. Then the imbibition stops at $\tilde{z}_{\rm min}$.

Finally, when $\tilde{z}_{\rm min}\ge 1$, or
\begin{equation}
\tilde{p}\ge \frac{1}{1-\alpha_{\rm F}}
\label{eq:T45}
\end{equation}
from Eq.~(\ref{eq:T44}), which reduces to the inequalities  (\ref{eq:T31}) and (\ref{eq:T37}), the driving pressure in Eq.~(\ref{eq:T30}) becomes always positive and the capillary flow to the end of capillary will be established. The forced imbibition is possible only when inequality ($\ref{eq:T31}$) or ($\ref{eq:T45}$) is satisfied.  Then, the steady capillary flow would be established, and the standard hydrodynamic approach could be used to study the dynamics and the time scale of capillary imbibition~\cite{Bell1906,Lucas1918,Washburn1921,Rideal1922,Bosanquet1923,Reyssat2008,Urteaga2013,Berli2014,Gorce2016}, which will be the subject of Sec.~\ref{sec:sec4}.

For the reverse direction, we introduce
\begin{eqnarray}
\tilde{p} &=& \frac{\bar{p}_{\rm ext}}{\left|\bar{p}_{\rm R,0}\right|},
\label{eq:T46} \\
\alpha_{\rm R} &=& \frac{H\tan\phi}{R_{\rm R,0}}=\frac{\alpha_{\rm F}}{1-\alpha_{\rm F}},
\label{eq:T47}\\
\Delta \tilde{{\rm g}}_{\rm R} &=& \frac{\Delta G_{\rm R}}{\pi R_{\rm R,0}^{2}\left|p_{\rm R,0}\right| H}.
\label{eq:T48}
\end{eqnarray}
Then, the free-energy landscape in Eq.~(\ref{eq:T34}) for the reverse direction is given by
\begin{equation}
\Delta \tilde{{\rm g}}_{\rm R}\left(\tilde{z}\right)=\left(1-\tilde{p}\right)\tilde{z}-\left(\tilde{p}-\frac{1}{2}\right)\alpha_{\rm R}\tilde{z}^{2}-\frac{1}{3}\tilde{p}\alpha_{\rm R}^{2}\tilde{z}^{3},
\label{eq:T49}
\end{equation}
where $\tilde{z}$ is defined in Eq.~(\ref{eq:T39}). The imbibition starts at $\tilde{z}=0$ when $\tilde{p}>1$, which corresponds to inequality (\ref{eq:T36}). The free-energy landscape has two extremums at $\alpha_{\rm R}\tilde{z}=-1$ and $\alpha_{\rm R}\tilde{z}=\left(1-\tilde{p}\right)/\tilde{p}$. The former belongs to the unphysical domain $\tilde{z}<0$. The latter corresponds to the free-energy maximum~\cite{Iwamatsu2020} at
\begin{equation}
\tilde{z}_{\rm max}=\frac{1-\tilde{p}}{\alpha_{\rm R}\tilde{p}},
\label{eq:T50}
\end{equation}
which occurs only when $\tilde{p}<1$.

\begin{figure}[htbp]
\begin{center}
\includegraphics[width=0.8\linewidth]{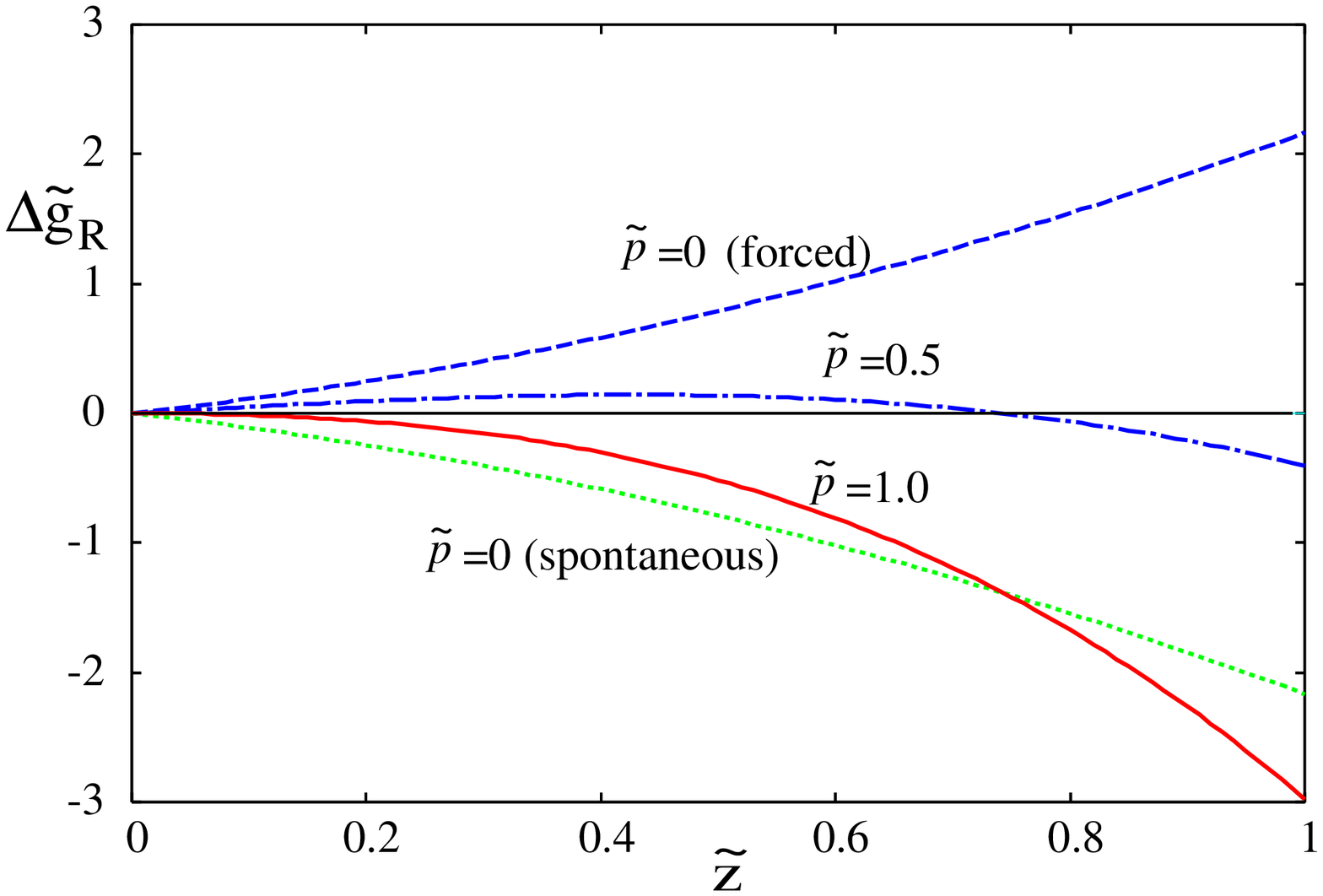}
\end{center}
\caption{
The free-energy landscape $\Delta \tilde{\rm g}_{\rm R}\left(\tilde{z}\right)$ in Eq.~(\ref{eq:T49}) for the reverse direction. The free-energy landscapes of the forced and the spontaneous imbibition with no applied pressure $\tilde{p}=0$ are also shown. The geometrical parameter is $\alpha_{\rm F}=0.7$ so that $\alpha_{\rm R}=2.33$. The free-energy landscape of the reverse direction exhibits maximum when $\tilde{p}\le 1$.  The forced imbibition starts at the inlet of the capillary when $\tilde{p}>1$ in Eq.~(\ref{eq:T36}). Then, the forced imbibition throughout the capillary is established.
}
\label{fig:7}
\end{figure}

Figure~\ref{fig:7} presents the free-energy landscape $\Delta \tilde{{\rm g}}_{\rm R}\left(\tilde{z}\right)$ for the reverse direction for various reduced pressures $\tilde{p}$. We also show the free-energy landscapes when $\tilde{p}=0$ and $\bar{p}_{\rm R,0}<0$ (forced imbibition) as well as when $\tilde{p}=0$ and $\bar{p}_{\rm R,0}>0$ (spontaneous imbibition) whose free energy (in original unit) is given by Eq.~(\ref{eq:T26}). The free-energy landscape shows maximum at $\tilde{z}_{\rm max}$ given by Eq.~(\ref{eq:T50}) when $\tilde{p}<1$ which moves toward $\tilde{z}=0$ as the pressure $\tilde{p}$ is increased to $\tilde{p}=1$.

This free-energy maximum indicates a negative driving pressure from Eq.~({\ref{eq:T30}) for $\tilde{z}<\tilde{z}_{\rm max}$, and the imbibition is prohibited.  The maximum reaches at $\tilde{z}_{\rm max}=0$ when $\tilde{p}=1$ ($\bar{p}_{\rm ext}=\left|\bar{p}_{\rm F,0}\right|$). Then, the driving pressure in Eq.~(\ref{eq:T30}) becomes always positive along the capillary and, the forced imbibition starts at the inlet of the capillary ($\tilde{z}=0$). Therefore, the capillary flow through whole capillary will be established as soon as
\begin{equation}
\tilde{p}\ge 1,
\label{eq:T51}
\end{equation}
which is identical to the inequalities (\ref{eq:T32}) and (\ref{eq:T36}).

In summary, the forced imbibition for the forward direction becomes possible only when the inequality (\ref{eq:T31}) or (\ref{eq:T45}) are satisfied, and that for the reverse direction becomes possible only when the inequality (\ref{eq:T32}) or (\ref{eq:T51}) is satisfied. The ratio of the minimum pressure $\left(p_{\rm ext}\right)_{\rm Forward}$ to $\left(p_{\rm ext}\right)_{\rm Reverse}$ which satisfy those inequities becomes
\begin{eqnarray}
\frac{\left(p_{\rm ext}\right)_{\rm Forward}}{\left(p_{\rm ext}\right)_{\rm Reverse}} &=& \frac{\left(1+\tan\phi\nu_{\rm R}\left(\theta_{\rm Y},\phi\right)\right)\Pi_{\rm F}\left(\theta_{\rm Y},\phi\right)}{\left(1-\tan\phi\nu_{\rm F}\left(\theta_{\rm Y},\phi\right)\right)\Pi_{\rm R}\left(\theta_{\rm Y},\phi\right)}
\nonumber \\
&\simeq&  \frac{\Pi_{\rm F}\left(\theta_{\rm Y},\phi\right)}{\Pi_{\rm R}\left(\theta_{\rm Y},\phi\right)}<1
\label{eq:T52}
\end{eqnarray}
from Fig.~\ref{fig:4} when Young's contact angle $\theta_{\rm Y}$ and the tilt angle $\phi$ belong to region III of the forced imbibition (Fig.~\ref{fig:5}).  By selecting the external pressure $p_{\rm ext}$ in the interval $\left(p_{\rm ext}\right)_{\rm Reverse}>p_{\rm ext}>\left(p_{\rm ext}\right)_{\rm Forward}$, only the forced imbibition for the forward direction will occur. Therefore, the liquid diode of one-way transport for the forward direction can be realized by the forced imbibition as well.

\section{\label{sec:sec4}Hydrodynamics and scaling rule of the capillary flow}

Once the steady flow is established, we can study the time scale of imbibition. Hagen-Poiseuille's law states~\cite{Landau1987,Staples2002,Young2004,Reyssat2008} that the volume flow rate $dV_{i}/dt$ is a constant and is given by
\begin{equation}
\frac{dV_{i}}{dt}=\frac{\pi\left[R_{i}(z)\right]^{4}}{8\eta}\frac{dp_{i}}{dz},
\label{eq:T53}
\end{equation}
for $i=$F and $i=$R, which can be integrate to give
\begin{equation}
\Delta p_{i}=\frac{8\eta}{\pi}\frac{dV_{i}}{dt}\int_{0}^{z}\frac{dz'}{\left[R_{i}(z')\right]^{4}},
\label{eq:T54}
\end{equation}
where $z$ is the position of meniscus.  For the capillaries with the circular cross section, the volume flow rate can be written as
\begin{equation}
\frac{dV_{i}}{dt}=\pi\left[R_{i}(z)\right]^{2}\frac{dz}{dt}.
\label{eq:T55}
\end{equation}
By identifying the pressure $\Delta p_{i}$ by Eq.~(\ref{eq:T30}) and replacing the volume frow late $dV_{i}/dt$ by Eq.~(\ref{eq:T55}), Eq.~(\ref{eq:T54}) is written as
\begin{equation}
p_{i,{\rm L}}(z)+p_{\rm ext}=8\eta\left[R_{i}(z)\right]^{2}\frac{dz}{dt}\int_{0}^{z}\frac{dz'}{\left[R_{i}(z')\right]^{4}}.
\label{eq:T56}
\end{equation}
Therefore, the time evolution of the meniscus follows
\begin{equation}
\frac{dz}{dt}=\frac{p_{i,{\rm L}}(z)+p_{\rm ext}}{8\eta\left[R_{i}(z)\right]^{2}\int_{0}^{z}\frac{dz'}{\left[R_{i}(z')\right]^{4}}}.
\label{eq:T57}
\end{equation}
This equation is valid as far as the capillary has a circular cross section~\cite{Staples2002,Reyssat2008}.  Now, we consider the special case of conical capillaries in Eqs.~(\ref{eq:T3}) and (\ref{eq:T4}).

For the spontaneous imbibition without the external pressure ($p_{\rm ext}=0$) in region I (Fig.~\ref{fig:5}), we recover the standard formula~\cite{Staples2002,Reyssat2008,Urteaga2013,Gorce2016,Singh2020}
\begin{equation}
\frac{dz}{dt}=\frac{c_{i}\gamma_{\rm lv}}{8\eta \left[R_{i}(z)\right]^{3}\int_{0}^{z}\frac{dz'}{\left[R_{i}(z')\right]^{4}}},
\label{eq:T58}
\end{equation}
where the constant $c_{i}$ is given by
\begin{equation}
c_{i}=\frac{2\Pi_{i}\left(\theta_{\rm Y},\phi\right)}{1\mp \tan\phi\nu_{i}\left(\theta_{\rm Y},\phi\right)}
\label{eq:T59}
\end{equation}
from Eq.~(\ref{eq:T16}).  Many authors used a simpler and intuitive expression $c_{i}=\cos\left(\theta_{Y}\mp \phi\right)$ from Eq.~(\ref{eq:T24})~\cite{Staples2002,Reyssat2008,Gorce2016,Singh2020} or $c_{i}=\cos\left(\theta_{Y}\right)$~\cite{Urteaga2013,Gorce2016}, where the upper $-$ in $\mp$ applies to the forward direction ($i=$F) and the lower $+$ applies to the reverse direction ($i$=R).  The spontaneous imbibition occurs for both the forward and the reverse direction in the regions I of Fig.~\ref{fig:5} where $\Pi_{i}\left(\theta_{\rm Y},\phi\right)>0$ or $c_{i}>0$.

For the conical capillaries in Eqs.~(\ref{eq:T3}) and (\ref{eq:T4}), the integral in Eq.~(\ref{eq:T58}) can be evaluated and we obtain~\cite{Reyssat2008}
\begin{equation}
\left[\left(1\mp Z_{i}\right)^{3}-1\right]\frac{dZ_{i}}{dT_{i}}=\mp\frac{3}{8},
\label{eq:T60}
\end{equation}
where the scaled length $Z_{i}$ and the scaled time $T_{i}$ are defined by
\begin{eqnarray}
Z_{i} &= \frac{\tan\phi z}{R_{i,0}},
\label{eq:T61} \\
T_{i} &= \frac{c_{i}\gamma_{\rm lv}\tan^{2}\phi}{\eta R_{i,0}}t.
\label{eq:T62}
\end{eqnarray}
Though Eq.~(\ref{eq:T60}) can be analytically solved~\cite{Reyssat2008}, we consider the asymptotic form. Equation.~(\ref{eq:T60}) can be simplified into
\begin{equation}
3Z_{i}\frac{dZ_{i}}{dT_{i}}=\frac{3}{8},
\label{eq:T63}
\end{equation}
at short distance $Z_{i}\ll 1$. Then, we recover the well-known Lucas-Washburn scaling rule
\begin{equation}
Z_{i}=\frac{1}{2}T_{i}^{1/2}\propto T_{i}^{1/2}.
\label{eq:T64}
\end{equation}
At long distances $Z_{\rm R}\gg 1$, which is realizable only for the reverse direction $i=$R since 
\begin{equation}
Z_{\rm F} = \frac{\tan\phi z}{R_{\rm{F},0}}<\frac{\tan\phi H}{R_{\rm{F},0}}=\frac{R_{\rm{F},0}-R_{\rm{R},0}}{R_{\rm{F},0}}<1,
\label{eq:T65}
\end{equation}
Equation~(\ref{eq:T60}) can be simplified into
\begin{equation}
Z_{i}^{3}\frac{dZ_{i}}{dT_{i}}=\frac{3}{8},
\label{eq:T66}
\end{equation}
and we recover the result of Reyssat et al.~\cite{Reyssat2008},
\begin{equation}
Z_{\rm R}=\left(\frac{3}{2}T_{\rm R}\right)^{1/4}\propto T_{\rm R}^{1/4}.
\label{eq:T67}
\end{equation}
For long distances $Z_{\rm R}\gg 1$, therefore, the capillary imbibition for the forward direction following the scaling rule (\ref{eq:T64}) is faster than that for the reverse direction following the scaling rule (\ref{eq:T67}).  The Lucas-Washburn scaling rule (\ref{eq:T64}) applies to the forward direction of liquid diode in region II of Fig.~\ref{fig:5} as well.

For short distances $Z_{i}\ll 1$, the completion time $t_{\rm F}$ (forward) and $t_{\rm R}$ (reverse) of capillary imbibition when the liquid meniscus reaches the outlet of the capillary are obtained from Eqs.~(\ref{eq:T62}) and (\ref{eq:T61}) at $z=H$, and using the scaling rule in Eq.~(\ref{eq:T64}).  They satisfy
\begin{equation}
\frac{t_{\rm F}}{t_{\rm R}}=\frac{c_{\rm R}R_{\rm F,0}T_{\rm F}}{c_{\rm F}R_{\rm R,0}T_{\rm R}} =\frac{c_{\rm R}R_{\rm F,0}Z_{\rm F}^{2}}{c_{\rm F}R_{\rm R,0}Z_{\rm R}^{2}}= \frac{c_{\rm R}R_{{\rm R},0}}{c_{\rm F}R_{{\rm F},0}}\sim \frac{R_{{\rm R},0}}{R_{{\rm F},0}}<1.
\label{eq:T68}
\end{equation}
Note that $c_{\rm R}\sim\Pi_{\rm R}<c_{\rm F}\sim\Pi_{\rm F}$ in region I (Fig.~\ref{fig:5}), where both $\Pi_{\rm R}$ and $\Pi_{\rm F}$ are positive (Fig.~\ref{fig:4}).
The completion time  $t_{\rm F}$ will be shorter than $t_{\rm R}$.  A more empirical formula for the whole range of capillary length:
\begin{equation}
\frac{t_{\rm F}}{t_{\rm R}}= \left(\frac{R_{{\rm R},0}}{R_{{\rm F},0}}\right)^{7/3}<1
\label{eq:T69}
\end{equation}
was proposed by Urteaga et al.~\cite{Urteaga2013} for nano-capillaries by numerically fitting the full analytical formula for Eq.~(\ref{eq:T60})~\cite{Reyssat2008}.  Again, the time scale $t_{\rm F}$ is shorter than $t_{\rm R}$.  Therefore, the imbibition for the forward direction is faster than the reverse direction~\cite{Urteaga2013,Berli2014,Gorce2016,Singh2020}.

For strong forced imbibition with $p_{\rm ext}\gg p_{i,{\rm L}}(z)$ in region III of Fig.~\ref{fig:5}, Eq.~(\ref{eq:T57}) is written as
\begin{equation}
\frac{dz}{dt}=\frac{c}{8\eta\left[R_{i}(z)\right]^{2}\int_{0}^{z}\frac{dz'}{\left[R_{i}(z')\right]^{4}}},
\label{eq:T70}
\end{equation}
with
\begin{equation}
c=p_{\rm ext},
\label{eq:T71}
\end{equation}
which leads to
\begin{equation}
\left[\left(1\mp Z_{i}\right)^{3}-1\right]\frac{dZ_{i}}{dT}=\mp\frac{3}{8}\left(1\mp Z_{i}\right),
\label{eq:T72}
\end{equation}
instead of Eq.~(\ref{eq:T60}), where the non-dimensional time $T$ is re-defined by
\begin{equation}
T=\frac{c\tan^{2}\phi}{\eta}t,
\label{eq:T73}
\end{equation}
which leads to the Lucas-Washburn scaling rule (\ref{eq:T64}) again:
\begin{equation}
Z_{i}\propto T^{1/2}
\label{eq:T74}
\end{equation}
at short distances $Z_{i}\ll 1$.  At long distances $Z_{\rm R}\gg 1$ for the reverse direction, we find
\begin{equation}
Z_{\rm R}=\left(\frac{9}{8}T\right)^{1/3}\propto T^{1/3},
\label{eq:T75}
\end{equation}
instead of Eq.~(\ref{eq:T67}).

The completion time $t_{\rm F}$ (forward) and $t_{\rm R}$ (reverse) are now obtained from Eqs.~(\ref{eq:T73}) and (\ref{eq:T61}) at $z=H$, and using the scaling rule in Eq.~(\ref{eq:T74}).  They satisfy an inequality,
\begin{equation}
\frac{t_{\rm F}}{t_{\rm R}}=\frac{T_{\rm F}}{T_{\rm R}} =\left(\frac{Z_{\rm F}}{Z_{\rm R}}\right)^{2}=\left(\frac{R_{{\rm R},0}}{R_{{\rm F},0}}\right)^{2}<1,
\label{eq:T76}
\end{equation}
similar to Eq.~(\ref{eq:T68}) at short distances $Z_{i}\ll 1$.  Again, the time scale $t_{\rm F}$ is shorter than $t_{\rm R}$: The forced imbibition for the forward direction is faster than the reverse direction.

\section{\label{sec:sec5} Conclusion}

In this study, we considered the spontaneous and the forced imbibition of liquid into a truncated conical capillary as a simplest model to study the effect of geometrical gradients and to assess the possibility of liquid diode by conical capillary tubes. We inferred that the conical capillary with converging or narrowing radius functions as the forward direction of the diode, whereas that with diverging or widening radius functions as the reverse direction. The critical contact angle for the onset of spontaneous imbibition of the former and latter belong to the hydrophobic and hydrophilic region, respectively, and they are determined from the tilt angle of the capillary wall. By selecting Young's contact angle for the capillary between two critical contact angles, only a forward direction of spontaneous imbibition with zero external pressure is realized.

Even when Young's contact angle is larger than two critical contact angles, the forced imbibition is possible. Furthermore, the forced imbibition solely in the forward direction can be realized by tuning the applied pressure. Therefore, the conical capillary still acts as a liquid diode for the forced imbibition.

Finally, we considered the dynamics of imbibition using the scaling rule of imbibition~\cite{Bell1906,Lucas1918,Washburn1921,Rideal1922,Bosanquet1923,Reyssat2008,Urteaga2013,Berli2014,Gorce2016} derived from the Hagen-Poiseuille law of steady flow.  We found that both the spontaneous and the forced capillary flows for the forward direction are faster than that for the reverse direction.   These findings would be beneficial in elucidating and developing functioning micro- and nano-capillaries of conical shapes.  Of course, to understand the dynamics, in particular, the micro- and nano-scale dynamics, the slip-length at the wall~\cite{Bocquet2010,Tran-Duc2019} could be important. Various effects, such as the friction at the contact line~\cite{Fernandez-Toledano2021}, and the viscous dissipation~\cite{deGennes1985} on the dynamic contact angle, the line tension at the inclined wall of conical capillary~\cite{Jensen1999,Grosu2014}, and the contact line pinning by roughness~\cite{Chow1998}, may also influence the imbibition.

\section*{Author Declaration}
\subsection*{Conflict of interest}
The author declares no conflict of interest.


\bibliography{PhysFluiV4X}

\end{document}